\documentclass[11pt]{article}

\usepackage[utf8]{inputenc}
\usepackage[T1]{fontenc}
\usepackage[a4paper,margin=2cm]{geometry}

\usepackage{graphicx}
\usepackage{tabularx}
\usepackage{booktabs}
\usepackage{tcolorbox}
\usepackage{makecell}
\usepackage{amsmath,amssymb}
\usepackage[numbers,sort&compress]{natbib}
\usepackage{hyperref}
\hypersetup{colorlinks=true,linkcolor=black,citecolor=black,urlcolor=blue}
\usepackage{enumitem}

\title{From Brain Models to Executable Digital Twins: \\ Execution Semantics and Neuro--Neuromorphic Systems}
\author{Alexandre Muzy\\
International Laboratory on Learning Systems (ILLS)\\
CNRS -- McGill -- ETS -- MILA -- Universit\'e Paris-Saclay\\
\texttt{alexandre.muzy@cnrs.fr}}
\date{}

\begin{document}
\maketitle

\section*{Abstract}
Brain digital twins aim to provide faithful, individualized computational representations of brains as dynamical systems, enabling mechanistic understanding and supporting prediction of clinical interventions. Yet current approaches remain fragmented across data pipelines, model classes, temporal scales, and computing platforms, which prevents the preservation of execution semantics across the end-to-end workflow. This survey introduces \emph{physically constrained executability} as a unifying perspective for comparing approaches at the level of execution: whether an execution state is persistent, which events are permitted to update it (simulation, measurement, actuation), and how strongly execution is temporally and causally coupled to neurobiological dynamics. Building on modeling and simulation theory, I propose a taxonomy of execution regimes ranging from isolated offline models to coordinated co-simulation, to continuously executing digital twins sustained by online data assimilation, and ultimately to neuro--neuromorphic physical systems in which biological and computational dynamics are co-executed under shared physical constraints. The executability concept clarifies why accuracy alone is insufficient, and motivates an agenda centered on semantic interoperability, hybrid-time correctness, evaluation protocols, scalable reproducible workflows, and safe closed-loop validation. This survey adopts a systems and runtime-oriented perspective, enabling comparison of heterogeneous approaches based on their execution semantics rather than on model form or application domain alone.

\section{Introduction}

Brain digital twins are increasingly used by scientists to better understand the functioning of individual brains, and by clinicians to predict the impact of interventions. They aim to provide faithful digital representations of individual brains conceived as dynamic systems. To this end, they seek to link multiscale brain data with models from neuroscience \citep{ref4,ref44,Xiong2023,Fekonja2024}.

In this context, large-scale international initiatives such as the European Human Brain Project (HBP) \citep{ref2,ref3,Amunts2024}  and its successor infrastructure European Brain Research InfrastructureS (EBRAINS) aim at integrating brain data, models and computing technologies, to achieve reproducible and multiscale computational neuroscience. 

These developments build, implicitly and separately, on long-standing foundations from computational modeling and simulation theory \citep{ref178}. In this tradition, a model is not only defined by its static mathematical functions but also by its simulation, i.e. the series of computations of its state transitions over time, its simulated trajectory. The model's internal state transitions of a brain digital twin then remain consistent from data acquisition to the simulation execution on a computer. 

In practice, however, computations in a digital twin are rarely specified formally. Current brain digital twin approaches remain fragmented across data pipelines, model classes, temporal scales, and computational platforms. This fragmentation can be understood as a failure to preserve \textit{execution semantics} across the end-to-end brain digital twin pipeline. In practice, the causal ordering, timing, and duration of state transitions are not maintained consistently across data acquisition, interaction, modeling, simulation, and computation, relative to the dynamics of the underlying neurobiological system.

Here, \textit{execution} refers to the realization of state transitions on a given substrate, which may be digital (computers and/or chips), physical (the brain), or hybrid, when biological and computational substrates are coupled in a closed loop. Importantly, execution on a given substrate is not abstract: it is necessarily subject to the physical properties and limitations of that substrate.

In this survey, I introduce \textit{physically constrained executability} as a framework for designing brain digital twins in which execution semantics are explicitly preserved under substrate-dependent constraints. In the case of the brain, physical constraints include, for example, the maximum firing rate of neurons \citep{mascart2022scalability} and the latency of synaptic communication \citep{gattepaille2025delay}. Under this framework, executability is \emph{physically constrained}: state transitions must remain consistent with substrate-dependent limits (measurement latencies, bandwidth, bounded update rates, and actuation constraints), so that the digital twin remains causally and temporally aligned with neurobiological dynamics throughout measurement and interaction.

\textit{From this perspective, the main limitation of current brain digital twin approaches lies not in the expressiveness of their models, but in the absence of explicit execution semantics capable of sustaining coherent state evolution across time, data, and interaction.}

\begin{figure}[ht!]
  \centering
  \includegraphics[width=0.95\linewidth]{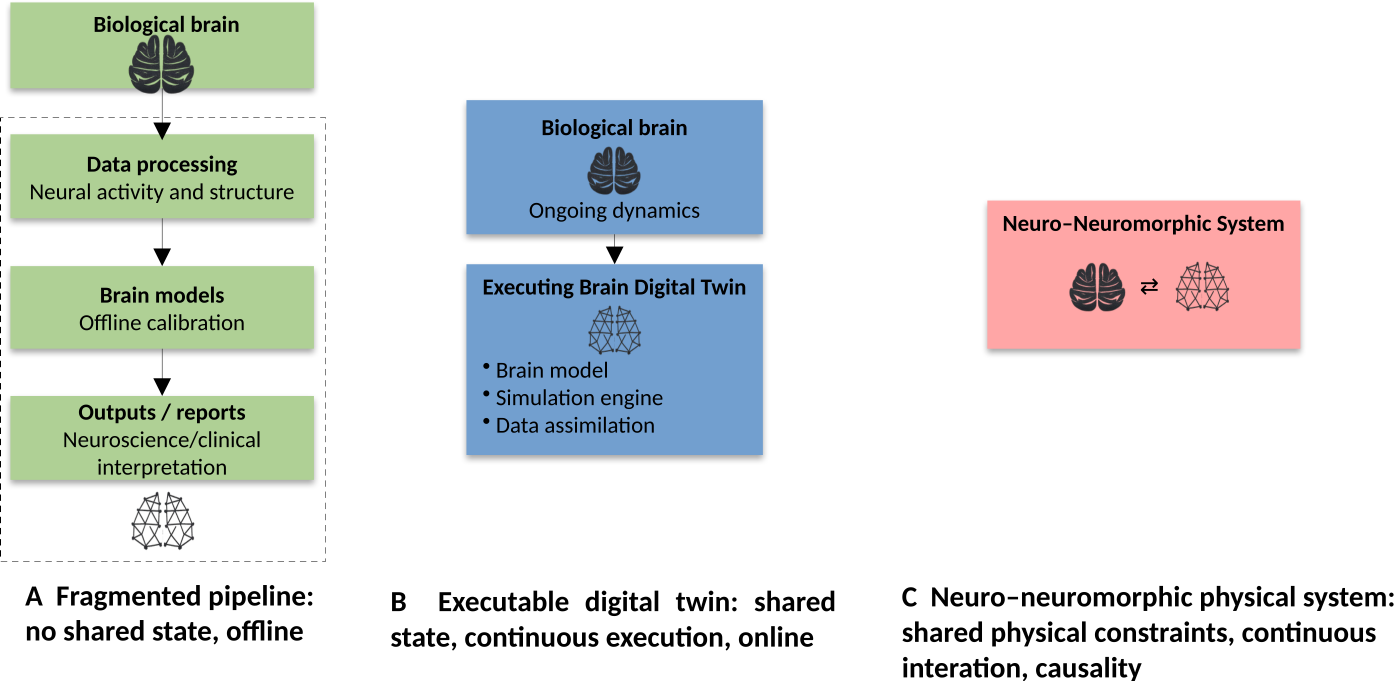}
  \caption{\textbf{From fragmented brain models to executable neuro--neuromorphic systems.}
  \textbf{(A)} Current brain digital twin approaches are fragmented across data processing, modeling, simulation, and neuroscientific/clinical interpretation. Each of these steps is typically organized as separate and offline pipelines. They are executed independently, with limited temporal coupling and no persistent feedback. This results in models that are biologically informed but not continuously executable.
  \textbf{(B)} Executable brain digital twins correspond to systems in which computational models are embedded within ongoing simulations coupled to data assimilation processes. Brain observations are integrated online to update model states and parameters during execution, enabling sustained temporal alignment between biological and digital dynamics. A key distinction at this stage is the presence of a persistent, shared execution state that is maintained over time and updated online through continuous observation and model evolution.
  \textbf{(C)} At the highest level of executability, neuro--neuromorphic systems co-execute biological and digital dynamics as a coupled system.}
  \label{fig:executability}
\end{figure}

Figure~\ref{fig:executability} provides a conceptual overview of the execution regimes discussed throughout this survey. Panel~(A) illustrates fragmented and episodic pipelines in which execution state is not preserved. Panel~(B) highlights the emergence of a persistent execution state continuously updated through \textit{data assimilation} \cite{james2024inferring}, which constitutes the minimal condition for executability. Here, data assimilation refers to the integration of empirical measurements during execution to update model states and parameters. Traditional calibration corresponds to an offline variant in which parameters are adjusted without maintaining a persistent execution state. Panel~(C) anticipates the additional physical execution constraints that arise when digital and biological dynamics are causally coupled, setting the stage for neuro--neuromorphic systems. 

Neuromorphic refers to a mode of execution in which computational processes operate under physical execution constraints comparable to those of neurophysical dynamics. A neuro--neuromorphic system is a closed-loop in which a brain digital twin is causally embedded via physical sensing and actuation interfaces. Biological dynamics, sensing, actuation, and digital computation are co-executed under shared physical execution constraints and explicit execution semantics.

\begin{figure}[t]
  \centering
  \includegraphics[width=0.95\linewidth]{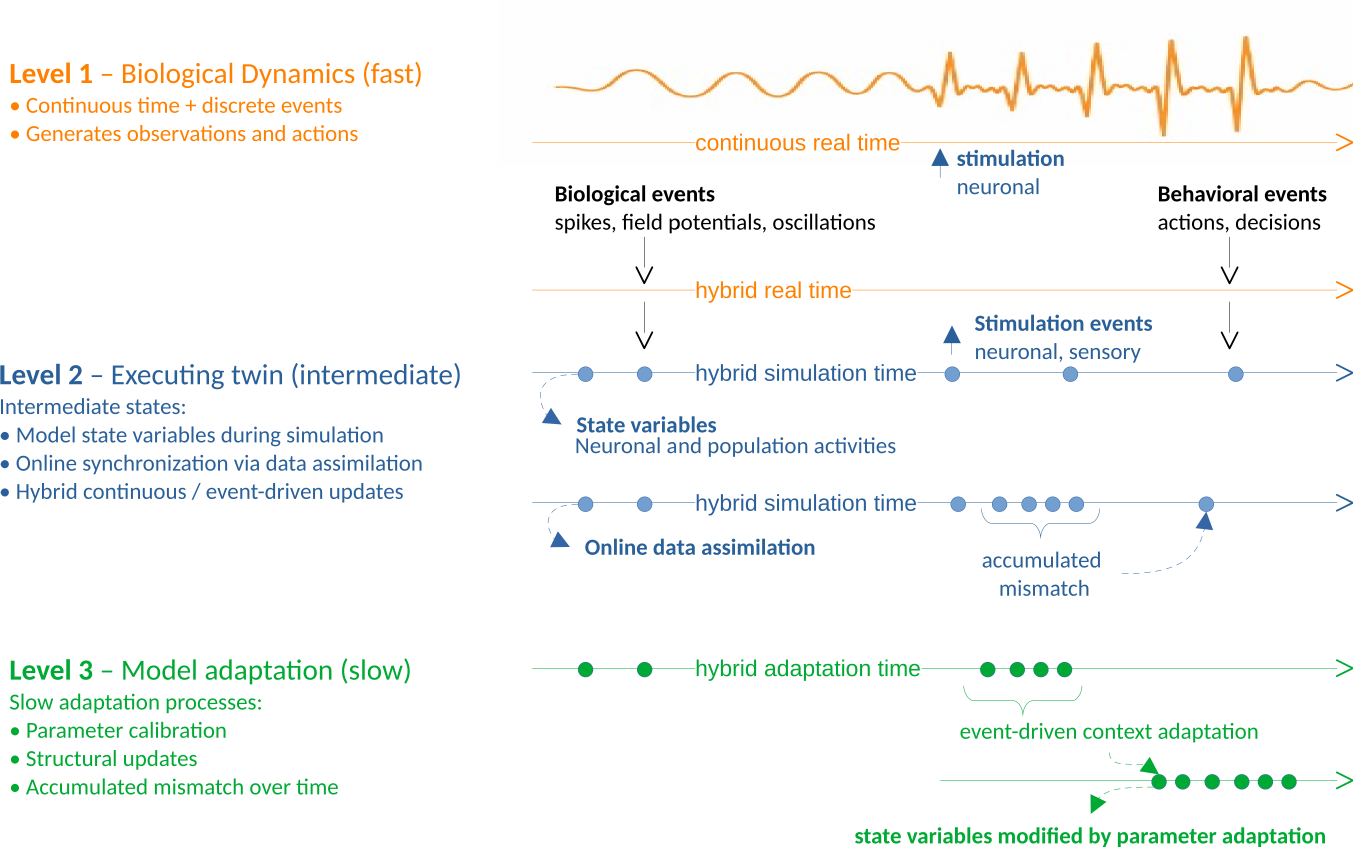}
 \caption{\textbf{Executability as state evolution across coupled time scales.}
Executable brain digital twins maintain a coherent execution state that evolves across
multiple, coupled temporal regimes.
\textbf{Level~1} corresponds to fast neurophysical dynamics of the biological brain,
including spiking activity, field potentials, oscillations, and behavior, which evolve
in continuous time and generate discrete events.
\textbf{Level~2} represents the executing state of the brain digital twin, defined by the
internal state variables of the computational model during simulation and inference.
This state evolves at an intermediate time scale and is continuously synchronized with
brain activity through online observations and data assimilation.
\textbf{Level~3} captures slow model adaptation processes, including parameter calibration
and structural updates, which integrate accumulated mismatches over time.
Together, these levels define a hybrid temporal semantics in which continuous dynamics,
discrete updates, and event-driven corrections coexist, enabling sustained execution
rather than episodic recalibration of the digital twin.}
  \label{fig:executability_timescales}
\end{figure}

Figure~\ref{fig:executability_timescales} further refines this perspective by representing executability as the evolution of a persistent execution state across coupled time scales. Executable brain digital twins maintain internal states that are continuously updated by heterogeneous events originating from neurobiological activity, measurements, computation, or action. As a result, executability inherently relies on hybrid temporal execution semantics, in which continuous dynamics coexist with discrete, event-driven updates across neurophysical time and simulation time.

\paragraph{Scope and methodology of the survey.}
This survey is concept-driven rather than exhaustive. The literature is organized
around execution-level questions that cut across neuroscience, modeling and simulation,
and neuromorphic computing, including the persistence of execution state, admissible
event types (simulation, measurement, actuation), and temporal and causal coupling
under physical constraints. References were selected to represent foundational theories,
widely used modeling pipelines, representative closed-loop systems, and execution
substrates exposing distinct runtime semantics. The objective is to provide a
comparative conceptual framework and evaluation criteria, rather than a catalog
of implementations.


Section~2 introduces a taxonomy of execution regimes, ranging from isolated models
to fully coupled neuro--neuromorphic systems. Section~3 focuses on data assimilation
and closed-loop personalization as execution modes that sustain persistent model
states over time. Section~4 examines how computational substrates constrain and
enable these regimes through their execution semantics and physical constraints.

\noindent
\begin{tcolorbox}[colback=gray!5,colframe=black,title=\textbf{Box 1 — Core system definitions}]
\small

\textbf{Brain Digital Twin (BDT):}
A brain digital twin is a computational brain model that maintains a persistent execution state that is continuously aligned with neurobiological observations acquired through sensing interfaces.\\[3pt]

\textbf{Physical execution constraints:}
Physical execution constraints arise from intrinsic brain dynamics and from sensing/actuation interfaces (e.g., bounded latency, bandwidth, noise, etc.).\\[3pt]

\textbf{Neuromorphic (execution mode):}
Neuromorphic refers to a mode of execution in which computational processes operate under physical execution constraints comparable to those of neurophysical dynamics.\\[3pt]

\textbf{Neuro--Neuromorphic System:}
A neuro--neuromorphic physical system is a closed-loop in which a brain digital twin is causally embedded via physical sensing and actuation interfaces. Biological dynamics, sensing, actuation, and digital computation are co-executed under shared physical execution constraints and explicit execution semantics.

\end{tcolorbox}

\section{Levels of Executability and Integration in Brain Digital Twins}

Brain digital twins (BDTs) cover a range of computational realizations, from offline models to physically coupled closed-loop systems \citep{ref18}. To structure this landscape, this section introduces a taxonomy of \textit{execution regimes} that characterizes how models are coupled and integrated, how execution time is handled, and how model states relate to measurements and \textit{actuation} on the brain during execution through causal relations.

Here, \textit{causality} denotes the directed influence by which events in one substrate trigger state transitions in another substrate according to the execution semantics of the system. Measurements correspond to causal influences from the brain to the digital twin, whereby neurobiological activity generates observation events that update the execution state of the model. 

By contrast, \textit{actuation} refers to causal influences from the digital twin to the brain mediated by physical interfaces, such as electrical or magnetic stimulation devices. Examples include closed-loop adaptive brain stimulation for epilepsy or movement disorders \citep{ref18,ref17}, connectomic deep brain stimulation \citep{ref54}, and model-informed neuromodulation strategies \citep{ref63}. In this context, actuation is distinct from clinical interventions performed by human operators, such as surgery or pharmacological treatment, which may be informed by digital twins but do not constitute actuation events within the execution loop.

Within this framework, differences between brain digital twin approaches can be traced back to how execution is organized and sustained, and in particular to how causality is instantiated when present. This includes how measurements and actuation events are generated, propagated, and integrated over time, as well as whether execution can be maintained coherently across model components and execution phases.

To make these differences comparable, each execution regime is defined using the same set of criteria: 
(i) the degree of \textit{temporal and causal coupling} to the underlying neurobiological process;
(ii) the \textit{types of events}---internal simulation events, measurement events, or actuation events---that are permitted to update the execution state;
and (iii) the presence or absence of a \textit{persistent execution state}, whose value is preserved across execution steps and execution phases, rather than being reinitialized at each simulation run.

Here, the term \emph{event} is used in the systems sense: a timestamped occurrence that triggers a state transition according to the execution semantics of the system, such as a simulation step, an measurement arrival, or an actuation command.

\begin{figure}[t]
  \centering
  \includegraphics[width=0.65\linewidth]{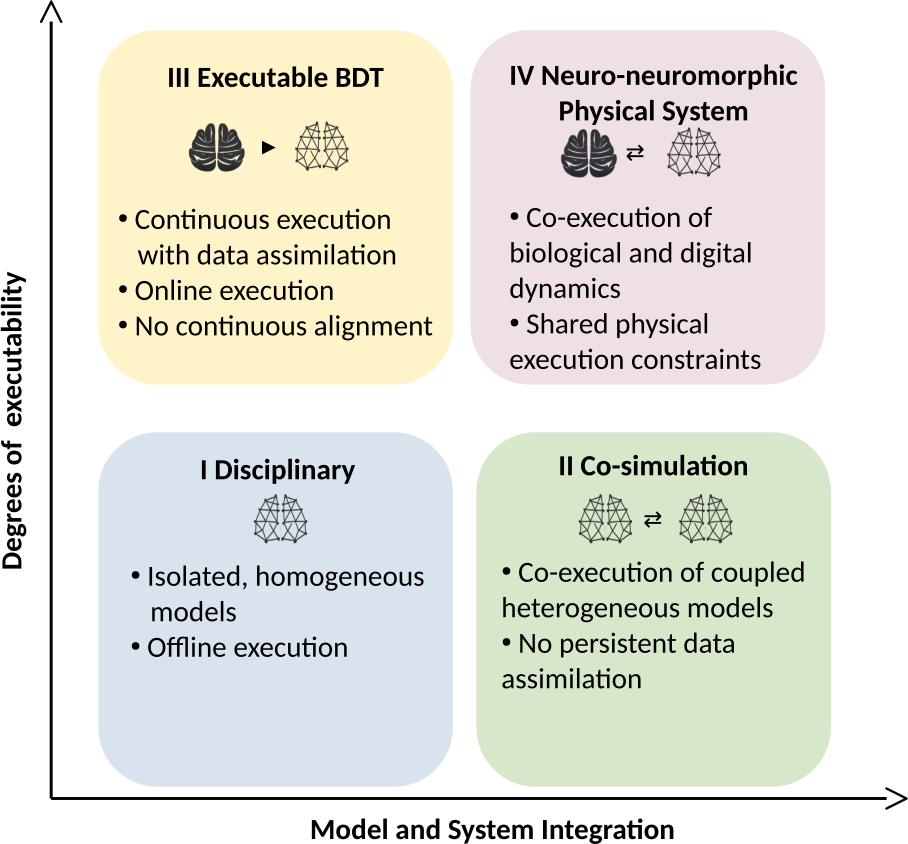}
  \caption{\textbf{Progressive executability of brain digital twins.}
The figure illustrates four execution regimes that characterize how brain digital twin approaches evolve from isolated computational models to fully embedded neuro--neuromorphic physical systems.
\textbf{Level I} corresponds to disciplinary models executed offline in isolation.
\textbf{Level II} captures co-simulation frameworks that enable the coordinated execution of coupled heterogeneous models without persistent data assimilation.
\textbf{Level III} represents executable brain digital twins that sustain continuous execution through online data assimilation and state adaptation, while remaining observational.
\textbf{Level IV} denotes neuro--neuromorphic physical systems, in which biological and digital dynamics are co-executed under physically constrained execution.
Together, these levels reflect increasing degrees of executability, integration, and causality.}
  \label{fig:model_to_sim}
\end{figure}

Figure~\ref{fig:model_to_sim} provides a schematic overview of this progression, from isolated computational models to fully coupled neuro--neuromorphic physical systems. The different levels of progression are described hereafter.

Importantly, this progression should not be interpreted as a mere increase in model integration. Rather, it reflects increasing degrees of \textit{executability}: the ability of a system to sustain coherent execution semantics across time, data, and interaction. Model integration, as introduced at Level~II, constitutes a necessary but insufficient condition for executability.

\subsection{Level I — Disciplinary isolated models (non-executable)}

At Level~I, brain models are developed within isolated disciplinary frameworks such as biophysical modeling, statistical inference, or machine learning. They are evaluated through analytical methods or standalone numerical simulations \citep{ref1,ref2,ref46,ref47}.

Let us examine the criteria defining this execution regime. There is \textit{no temporal or causal coupling} to the underlying neurobiological process. Empirical measurements may inform model structure, parameters, or initial conditions exclusively through offline data assimilation, including calibration, or through \textit{post hoc} analysis. The model state is updated solely through \textit{internal simulation events} occurring within an isolated simulation run. There is \textit{no persistent execution state}: the state exists only transiently during execution and is systematically reinitialized at each run. As a result, execution is episodic and does not support continuity across time, observations, or interactions.

As a result, there is no sustained temporal or causal coupling between the model and the biological system. Level~I models therefore function as descriptive or explanatory tools rather than executable systems, corresponding to the fragmented pipelines illustrated in Figure~\ref{fig:executability}A.

\subsection{Level II — Co-simulation frameworks (controlled execution)}

At Level~II, co-simulation frameworks enable the coordinated execution of multiple heterogeneous models or simulators, such as spiking neural networks coupled with neural mass or whole-brain models \citep{ref5,ref6,ref21,ref22}. 

Let us examine the criteria defining this execution regime. As in Level~I, there is \textit{no temporal or causal coupling} to the underlying neurobiological process during execution. Empirical data may inform initialization or parameterization offline, but do not enter execution as events. The key progression relative to Level~I lies in the \textit{integration of heterogeneous models} through co-simulation frameworks. The execution state is updated exclusively by \textit{internal simulation events}, but these events may originate from multiple coupled model components that must be temporally coordinated. A \textit{persistent execution state exists during a simulation run}, allowing continuous state evolution and synchronization across coupled models. However, this state is not preserved across execution phases and is reinitialized between runs. As a result, Level~II systems achieve structural and temporal integration, but remain non-executable with respect to the brain.

Level~II systems are therefore runnable and temporally coordinated, but not continuously executable in the sense defined here. Execution remains episodic and decoupled from ongoing neurobiological dynamics, with adaptation occurring primarily through offline calibration \citep{ref178,ref105}.

\subsection{Level III — Executable brain digital twins (adaptive execution)}

At Level~III, brain digital twins become continuously executing computational systems. The defining feature of this regime is the presence of a \emph{persistent execution state} that is maintained and updated over time.

Level~III represents executable brain digital twins that sustain continuous execution through online data assimilation and state adaptation, while remaining observational, without causal influence on the underlying neurobiological process. 

Let us examine the criteria defining this execution regime. There is a \textit{uni-directional temporal and causal coupling} from the brain to the model, without physical actuation. State updates are driven both by \textit{internal simulation events} and by \textit{external measurement events} originating from the biological system. Incoming neurophysiological measurements are integrated during execution through data assimilation mechanisms (Figure~\ref{fig:executability_timescales}), encompassing online state estimation and parameter adaptation \citep{james2024inferring}. The execution state is \textit{persistent} across measurement events and continuously updated rather than reinitialized. Despite this increased executability, Level~III systems remain observational: causal influence flows from the brain to the digital twin, without physical actuation or intervention.

Level~III thus constitutes the first regime that satisfies the core criteria of executability: a persistent execution state, event-driven updates, and sustained temporal coherence with neurobiological dynamics (Figure~\ref{fig:executability}B).

\subsection{Level IV — Neuro--neuromorphic physical systems (co-execution)}

Level~IV represents the most stringent regime of executability. Biological dynamics, computation, and intervention are integrated into a single closed-loop system.

Let us examine the criteria defining this execution regime. There is a \textit{bidirectional temporal and causal coupling} between the model and the underlying neurobiological process. Causal influence flows from the brain to the digital twin through \textit{external measurement events} that update the execution state, and from the digital twin to the brain through \textit{external actuation events} that physically affect the neurobiological system. The execution state is \textit{persistent} across measurement and actuation events and continuously updated through closed-loop interaction. In this regime, execution is subject to explicit \textit{physical execution constraints} imposed by brain dynamics, such as latency, bandwidth, and actuation limits, which must be satisfied throughout execution.

Coarse, batch-oriented execution models become insufficient under these conditions. Level~IV systems therefore rely on execution paradigms that support event-driven computation and persistent state under physical coupling. Neuromorphic execution substrates provide a natural support for this regime, although the defining characteristic is not the hardware itself, but the explicit satisfaction of physical execution constraints.

\textit{This regime corresponds to neuro--neuromorphic physical systems (Figure~\ref{fig:executability}C), enabling adaptive neuromodulation, closed-loop experimentation, and sustained interaction with living neural dynamics \citep{ref17,ref18}. Level~IV thus represents the limiting case of executability considered in this survey.}

\emph{From this taxonomy, executability does not appear as a binary property but as a progressive acquisition of execution capabilities: brain digital twins differ in whether an execution state is persistent, which types of events are permitted to update it, and to what extent execution becomes temporally and causally coupled to underlying neurobiological dynamics.}

\section{Closed-loop Personalization: Data Assimilation and Intervention}

\begin{figure}[t]
  \centering
  \includegraphics[width=0.6\linewidth]{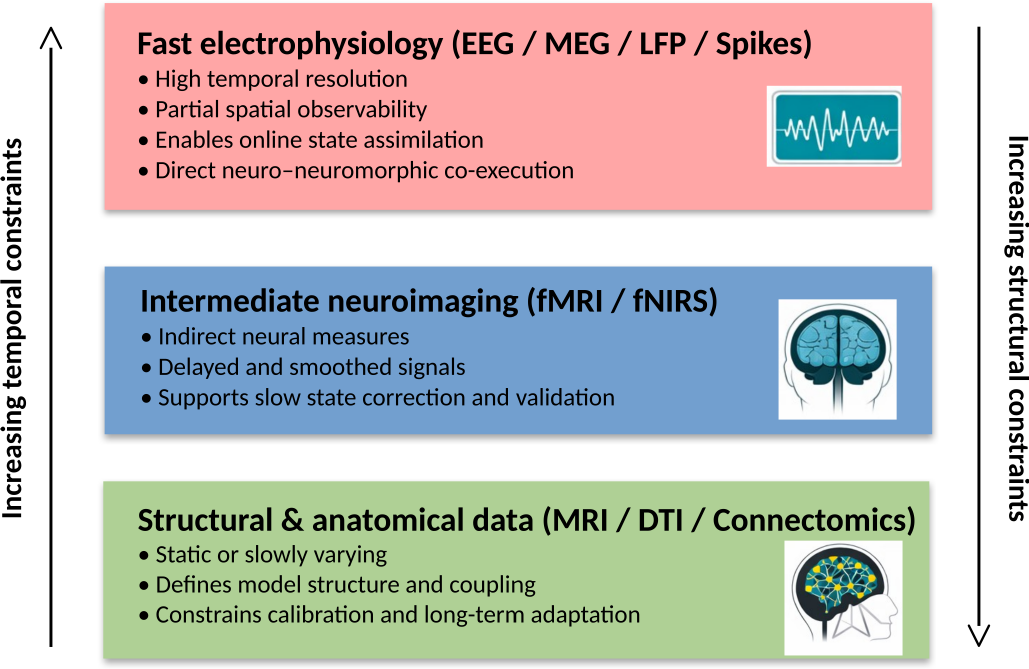}
  \caption{\textbf{Brain measurements supporting brain digital twins.}
  Multimodal measurements provide complementary access to brain structure,
  dynamics, and function across spatial and temporal scales.
  Structural imaging (MRI, diffusion MRI) constrains anatomy and connectivity. Functional imaging (fMRI, fNIRS, PET) captures large-scale activity and metabolism. Electrophysiological recordings (EEG, MEG, ECoG, SEEG) provide fast temporal sampling of neural dynamics; molecular and histological data refine cellular
  and receptor-level properties. Together, these measurements ground the
  construction, personalization, and validation of brain digital twins.}
  \label{fig:measurements}
\end{figure}

A core requirement for executable brain digital twins is their ability to continuously align model dynamics with empirical \textit{measurements}. This alignment relies on online state estimation and parameter adaptation techniques (broadly referred to here as data assimilation), which integrate multimodal neurophysiological and neuroimaging measurements during execution \citep{ref11,ref72,ref94,ref110} (cf. Figure~\ref{fig:measurements}). Data assimilation is used here in a broad sense, encompassing online state estimation and parameter adaptation, with offline calibration corresponding to its limiting non-executable case. Through such mechanisms, model states and parameters can be personalized over time, rather than fixed once through offline calibration \citep{ref12,ref25,ref105}.

In an executable brain digital twin, the execution state may include fast neural variables (e.g., population activity or oscillatory phase), latent physiological quantities (e.g., regional excitability or effective connectivity), probabilistic belief states over parameters, and slower contextual states related to clinical condition or behavior. These states are updated online by \textit{measurement-driven events}, enabling sustained temporal alignment between the digital twin and ongoing brain dynamics. The persistence of this execution state across measurement events is a defining feature of executability.

Crucially, continuous personalization through data assimilation introduces explicit \textit{execution constraints}. Measurements are not instantaneous abstractions but are acquired under physical and technical constraints, including sampling rates, latencies, spatial resolution, and noise characteristics. These constraints impose structural and temporal bounds on how measurement events can update the execution state, directly linking executability to the structural and temporal constraints illustrated in Figure~\ref{fig:executability}. In this sense, physically constrained execution requires that model updates remain temporally and causally consistent with the measurement processes of the brain.

Several approaches already incorporate closed-loop interaction and model-driven intervention, including adaptive brain stimulation and control strategies \citep{ref17,ref18,ref54,ref63}. Here, \emph{model-driven intervention} refers to actions on the neurobiological system that are causally generated from the evolving internal state of a computational model, rather than from fixed heuristics or direct signal thresholds. For example, inferred latent variables such as seizure likelihood, network synchrony, or excitation--inhibition balance may be used to adjust stimulation parameters in real time, including amplitude, timing, or spatial targeting.

Despite these advances, most existing closed-loop implementations remain task-specific and episodic, rather than fully executable brain digital twins. Model execution, data assimilation, and intervention are often implemented as loosely connected stages rather than as a single, temporally coherent execution regime. From an executability perspective, the central challenge is therefore not only to design accurate estimators or controllers, but to integrate simulation, data assimilation, and intervention within a unified execution process that preserves a persistent execution state across time.

Similar limitations arise in clinical neuroscience applications such as epilepsy modeling and neurosurgical planning, where individualized models inform prediction or intervention \citep{ref5,ref21,ref33,ref42}. While these approaches demonstrate the potential of personalized modeling, they typically rely on offline calibration and episodic simulation, leaving open the question of how closed-loop operation can be sustained as a continuous execution regime.

Addressing this challenge requires not only methodological advances, but also infrastructures capable of supporting persistent, constrained execution over long time scales. Large-scale international initiatives provide critical foundations in this respect. Programs such as the Human Brain Project and EBRAINS, the Human Connectome Project, and the U.S. BRAIN Initiative support long-term measurement acquisition, shared standards, and computational platforms that enable reproducible and scalable execution of brain models \citep{Markram2012HBP,Amunts2019EBRAINS,Glasser2016,Insel2013,Bargmann2014}. Under continuous execution, such standards no longer function solely as data-sharing conventions but become \textit{execution constraints} that condition how measurements can be integrated over time.

At a foundational level, FAIR (Findable, Accessible, Interoperable, Reusable) principles specify requirements on data and metadata that directly condition executability across time, platforms, and cohorts \citep{Wilkinson2016FAIR}. Building on these principles, standards such as the Brain Imaging Data Structure (BIDS) constrain acquisition parameters, timing information, and preprocessing assumptions in ways that directly affect how measurement events can be aligned with model states during execution \citep{ref81}. Together with standardized workflows, these frameworks enable execution states to be reproduced, compared, and re-evaluated across studies, rather than merely facilitating data exchange \citep{ref82,ref83,ref84,ref85}.

\emph{From an executability perspective, personalization is no longer a preprocessing step but an execution mode, in which model states and parameters are continuously updated under explicit temporal, structural, and measurement constraints.}

\begin{tcolorbox}[colback=gray!5,colframe=black,
title=\textbf{Box 2 — Measurements of brain structure and activity}]
\small
Brain digital twins rely on multimodal measurements that capture complementary
aspects of brain organization and dynamics across scales.\\[3pt]
\textbf{Acronyms:}
MRI, Magnetic Resonance Imaging; diffusion MRI, diffusion-weighted MRI; fMRI, functional MRI; fNIRS,
functional Near-InfRared Spectroscopy; PET, Positron Emission Tomography;
EEG, ElectroencEphaloGraphy; MEG, MagnetoEncephaloGraphy; ECoG,
electroCorticoGraphy; SEEG, StereoElectroEncephaloGraphy.\\[3pt]
\textbf{Structural imaging} (MRI, diffusion MRI) provides anatomical information
and estimates of white-matter connectivity.\\[3pt]
\textbf{Functional imaging} (fMRI, fNIRS, PET) measures large-scale brain activity
and metabolic or molecular processes.\\[3pt]
\textbf{Electrophysiology} (EEG, MEG, ECoG, SEEG) records neural signals with high
temporal resolution.\\[3pt]
\textbf{Microscopy, histology, and molecular data} inform cellular composition,
receptor distributions, and microcircuit organization.\\[3pt]
Together, these modalities provide the empirical basis for constructing,
personalizing, and validating brain digital twins.
\end{tcolorbox}

\section{Platforms and Execution: From Software to Neuromorphic Systems}

From a hardware viewpoint, computational platforms are not interchangeable implementation details: they impose specific scheduling models, synchronization mechanisms, and latency constraints that directly shape execution semantics.

Executable brain digital twins depend critically on the computational substrates on which they are executed. As argued in Sections~2 and~3, executability is not determined by model expressiveness alone, but by the ability of a system to sustain coherent state evolution, data assimilation, and interaction under explicit physical execution constraints. These requirements place fundamentally different demands on computational platforms.

In most current execution regimes, models are embedded within simulation and data assimilation loops supported by conventional or accelerated computing platforms, including \emph{central processing units} (CPUs), \emph{graphics processing units} (GPUs), \emph{tensor processing units} (TPUs), and \emph{neural processing units} (NPUs) \citep{ref6,ref14,ref53}. These platforms enable large-scale numerical simulation and data-driven inference, but rely on synchronous, clock-driven execution models based on discretized time steps, batch processing, and centralized scheduling \citep{muzy2017iterative,mascart2022scalability}. Such execution models are well suited for offline analysis and episodic execution, but provide limited support for continuous interaction with ongoing neurobiological dynamics.

As executability increases toward continuous adaptation, low-latency interaction, and sustained closed-loop operation, these execution models encounter intrinsic limitations. Clock-driven and batch-oriented architectures struggle to accommodate the asynchronous, irregular, and event-based update patterns that characterize neurobiological processes, even when massively parallel hardware is employed \citep{Brette2007Exact,Indiveri2015Neuromorphic}. This mismatch becomes critical at Level~IV, where sustained co-execution must satisfy explicit constraints imposed by continuous physical coupling.

Under physical coupling, execution constraints arise from interaction with a living neurobiological system and primarily concern temporal and causal consistency. These constraints include bounded latency, continuous temporal coupling, and causally ordered interaction between measurement and actuation events. Such constraints are not algorithmic design choices, but external conditions imposed by physics and neurophysiology, which determine how computation can be executed, synchronized, and coupled to biological dynamics during execution.

\begin{figure}[t!]
   \centering
   \includegraphics[width=0.7\linewidth]{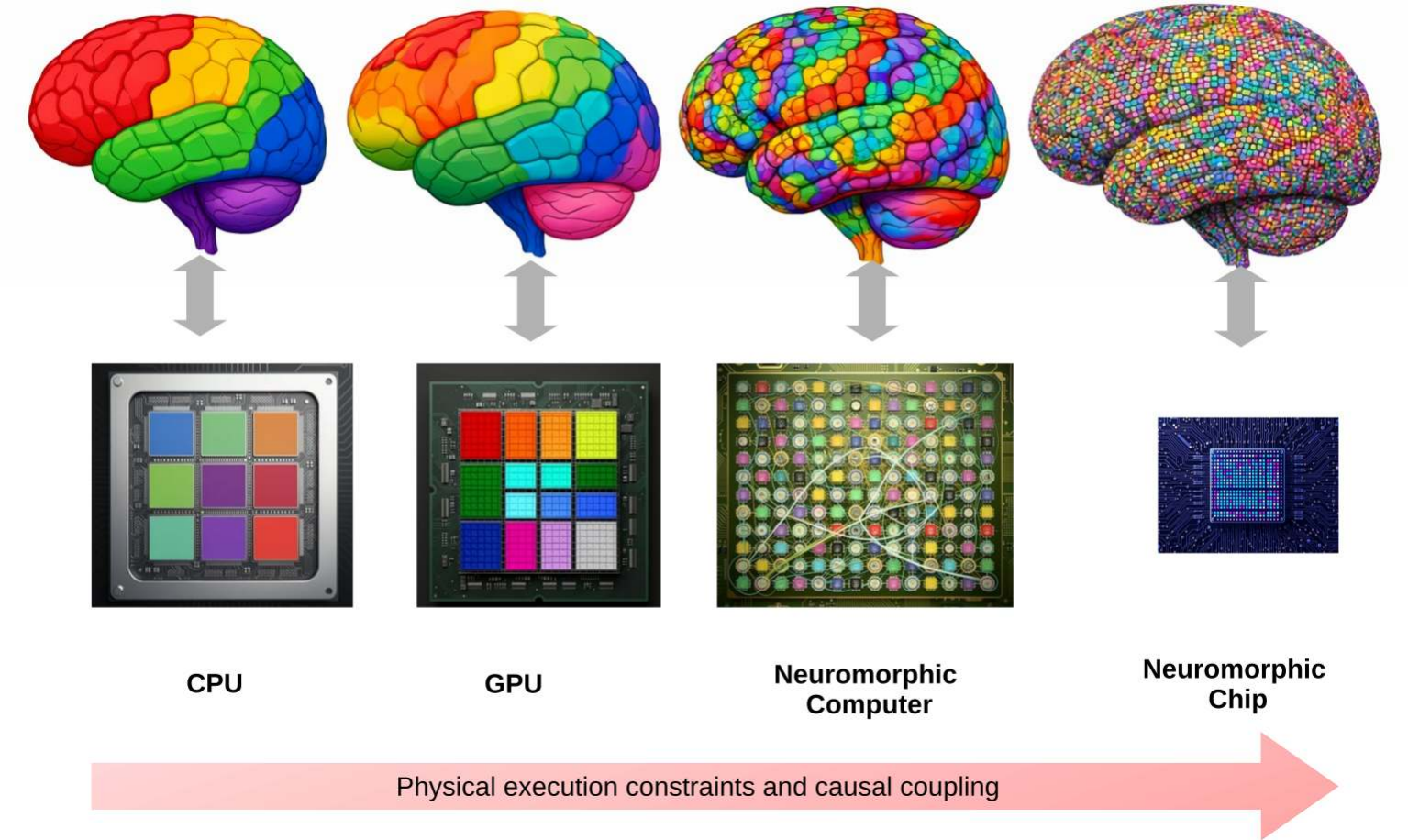}
   \caption{\textbf{Emergent execution granularity under physical execution constraints.}
This illustration compares major computational execution substrates with progressively finer representations of the same biological brain. From left to right, increasing physical execution constraints---such as bounded latency, causal coupling, and event-level interaction---impose increasingly fine-grained representations and execution units.
\textbf{CPU-based platforms} are associated with coarse-grained brain representations, depicted as large colored regions.
\textbf{GPU-based platforms} support finer-grained representations, illustrated by denser color patterns reflecting massively parallel processing units.
\textbf{Neuromorphic supercomputers} (e.g., SpiNNaker-like systems) correspond to distributed representations composed of interacting neuronal groups.
\textbf{Neuromorphic chips} operate at the finest execution granularity.
Overall, the figure highlights that representational and execution granularity do not arise as independent design choices, but emerge as a direct consequence of the physical constraints under which computation must be executed.}
\label{fig:execution}
\end{figure}

Under such constraints, the granularity at which computation is expressed becomes a central issue. In this survey, \emph{execution granularity} denotes the spatial, temporal, and causal resolution at which model states are represented, updated, and coupled to measurements and actions during execution. Preserving temporal and causal coherence can no longer rely on coarse, aggregated state updates or infrequent synchronization. Instead, computations must be expressed at increasingly fine spatial, temporal, and interaction granularity, such that individual events and local state changes can be causally aligned with neurophysical dynamics. Importantly, this granularity does not arise as an independent modeling choice, but as a necessary consequence of executability under physical execution constraints. Figure~\ref{fig:execution} illustrates how increasingly stringent constraints induce an emergent alignment between brain representations and the execution units provided by different computational substrates.

In this context, neuromorphic paradigms constitute a distinct \emph{execution regime} rather than an alternative modeling approach. Neuromorphic systems are characterized by spike-based, event-driven neuronal computation, local state updates, and reduced reliance on global synchronization. However, a critical distinction must be made between \emph{event-driven computation} and \emph{event-driven execution at the hardware level}. Executability under physical constraints requires not only that neurons compute in an event-driven manner, but that communication, scheduling, and synchronization across processing elements are themselves governed by events rather than by a global clock.

At the device level, neuromorphic chips such as Loihi, TrueNorth, or Akida implement spike-based, event-driven neuronal computation and support low-latency, energy-efficient processing suitable for embedded inference and local closed-loop interaction \citep{Merolla2014TrueNorth,Davies2018Loihi}. However, their execution remains partially structured by clocked control, time-multiplexing, or predefined scheduling mechanisms, limiting explicit control over system-level temporal and causal ordering.

At larger scales, neuromorphic supercomputers such as SpiNNaker or BrainScaleS extend event-driven principles to the level of distributed execution. These platforms support asynchronous communication, event-driven scheduling, and explicit handling of timing and causal propagation across thousands to millions of processing elements, enabling large-scale neural co-execution under tight temporal constraints \citep{Furber2014SpiNNaker,BrainScaleS2}.

In this regime, computation is no longer an external control layer applied to the brain, but becomes causally embedded in ongoing neurophysical activity. Neuro--neuromorphic physical systems thus represent the most stringent regime of executability discussed in this survey, in which biological and digital dynamics are co-executed under shared temporal, adaptive, and physical constraints \citep{Amunts2024,DeDomenico2025}.

\emph{From an executability standpoint, computational platforms are therefore not interchangeable implementation choices, but define distinct execution regimes according to the temporal, causal, and physical constraints they can sustain during continuous computation.}

\newpage
\begin{tcolorbox}[colback=gray!5,colframe=black,
title=\textbf{Box 3 — Execution substrates and constraints in brain digital twins}]
\small
Executable brain digital twins are constrained by the computational substrates on which they are executed, as these digital substrates determine the temporal, spatial, and causal granularity at which execution can be sustained under physical coupling.

\textbf{Conventional CPU-based platforms} rely on \emph{Central Processing Units (CPUs)}, which execute instructions sequentially or with limited core-based parallelism under a global clock. They support flexible numerical simulations and multiscale model composition in discretized time with global synchronization as shown in \citep{mascart2022scalability}. This makes them well suited for offline analysis, theoretical exploration, and episodic execution, but ill-suited for continuous interaction with ongoing biological dynamics.\\[3pt]
\textbf{Accelerated platforms} include \emph{Graphics Processing Units (GPUs)}, optimized for massively parallel numerical operations; \emph{Tensor Processing Units (TPUs)}, specialized for dense linear algebra in machine learning workloads; \emph{Neural Processing Units (NPUs)}, designed for efficient inference and embedded AI; and large-scale \emph{High-Performance Computing (HPC)} systems that aggregate thousands of such processors, including CPUs. These platforms enable large-scale and high-throughput simulation through clock-driven, synchronous execution, but remain fundamentally batch-oriented, limiting native support for asynchronous, event-driven, and continuously adaptive interaction.\\[3pt]
\textbf{Neuromorphic execution substrates} implement spike-based, event-driven neuronal computation with local state updates and reduced reliance on global clocks. Embedded neuromorphic chips (e.g., Loihi, TrueNorth, Akida) support low-latency and energy-efficient execution at the level of individual neurons and synapses, but retain partial clocked control at the hardware level. In contrast, large-scale neuromorphic supercomputers (e.g., SpiNNaker, BrainScaleS) implement event-driven execution across communication, scheduling, and synchronization, enabling distributed co-execution of neural dynamics under explicit temporal and causal constraints.

Rather than interchangeable implementation choices, these substrates define successive execution regimes by constraining the granularity, latency, and causal structure under which executable brain digital twins can operate.
\end{tcolorbox}

\section{Conclusion and Perspective: Toward Executable Neuro--Neuromorphic Systems}

This survey has traced a unifying trajectory from brain models to simulations, from simulations to executable systems, and from execution to closed-loop interaction with biological brains. The central message is that the main challenge of brain digital twins is no longer model construction \textit{per se}, but their operationalization under explicit computational, temporal, and physical execution constraints. From this perspective, \textit{physically constrained executability} emerges as the unifying computational concept that links heterogeneous model composition, hybrid-time simulation, continuous data assimilation, and ultimately neuro--neuromorphic co-execution, while preserving execution semantics across substrates.

Progress toward clinically and scientifically relevant brain digital twins therefore hinges on a focused set of open problems: (i) semantic interoperability across data and model interfaces; (ii) correctness under hybrid and real-time temporal regimes; (iii) identifiability and uncertainty quantification under partial and temporally constrained measurements; (iv) benchmarking and evaluation protocols for executable systems; (v) Cohort-level scalability and population validation, which pose execution-level challenges, as executable brain digital twins must preserve their execution semantics consistently across individuals, datasets, and execution contexts, rather than relying on subject-specific \textit{ad hoc} execution regimes; (vi) reproducible workflows, standards, and containerized execution; and (vii) closed-loop validation under adaptive and physically constrained intervention. Together, these challenges highlight that executability provides not only a design principle, but also a basis for systematic evaluation and comparison of brain digital twin approaches in terms of their execution semantics and constraints.

The translation of executable brain digital twins to large-scale and clinical settings further requires careful consideration of reproducibility, cohort-level validation, and scalability \citep{ref51}. Such efforts critically depend on standardized neuroimaging formats and processing pipelines, which provide the infrastructure necessary for interoperable and reproducible execution workflows \citep{ref81,ref82,ref83,ref84,ref85}. At a global level, major international initiatives continue to play a central role in structuring data sharing, computational platforms, and integrative research agendas \citep{ref2,ref6,Amunts2024}. Together, these developments point toward a unifying computational perspective in which executable, closed-loop neuro--neuromorphic systems emerge as a natural extension of current brain digital twin research \citep{DeDomenico2025}.

\emph{Taken together, these considerations suggest that progress in brain digital twins depends less on isolated modeling advances than on the explicit design of physically constrained executable systems capable of preserving execution semantics and sustaining coherent state evolution under continuous temporal, causal, and physical constraints.}

\bibliographystyle{unsrt}
\bibliography{references}

\end{document}